\documentclass[aps,prb,superscriptaddress,showpacs,twocolumn]{revtex4-1}
\pdfoutput=1
\usepackage{graphicx}
\usepackage{amsmath}
\usepackage{bbm}
\usepackage{xspace}
\usepackage{color}
\usepackage{sidecap}
\usepackage{hyperref}

\usepackage[normalem]{ulem}

\DeclareMathOperator{\re}{\mbox{Re}}

\newcommand{\se}{Sec.\@\xspace}

\newcommand{\eqw}[1]{(\ref{#1})}
\newcommand{\eq}[1]{Eq.\thinspace{}(\ref{#1})}
\newcommand{\eqq}[2]{Eqs.\thinspace{}(\ref{#1}) and (\ref{#2})}

\newcommand{\fig}[1]{Fig.\thinspace{}\ref{#1}}

\newcommand{\figgg}[3]{Figs. \ref{#1}, \ref{#2}, and \ref{#3}}
\newcommand{\fc}[1]{({#1})}
\newcommand{\figc}[2]{Fig.\thinspace{}\ref{#1}\thinspace{}\fc{#2}}

\renewcommand{\d}{\ket{\downarrow}}
\renewcommand{\u}{\ket{\uparrow}}

\newcommand{\be}{\begin{equation}}
\newcommand{\ee}{\end{equation}}
\newcommand{\bea}{\begin{eqnarray}}
\newcommand{\eea}{\end{eqnarray}}

\newcommand{\la}{\langle}
\newcommand{\ra}{\rangle}

\renewcommand{\phi}{\varphi}
\renewcommand{\epsilon}{\varepsilon}

\def\nn{\nonumber\\}

\def\bra#1{\mathinner{\langle{#1}|}}
\def\ket#1{\mathinner{|{#1}\rangle}}

\begin{document}

\title{Time dependent impurity in ultracold fermions: orthogonality catastrophe and beyond}

\author{Michael Knap}
\affiliation{Institute of Theoretical and Computational Physics, Graz University of Technology, 8010 Graz, Austria}
\affiliation{Department of Physics, Harvard University, Cambridge MA 02138, USA}

\author{Aditya Shashi}
\affiliation{Department of Physics and Astronomy, Rice University, Houston, Texas 77005, USA}

\author{Yusuke Nishida}
\affiliation{Theoretical Division, Los Alamos National Laboratory,
Los Alamos, New Mexico 87545, USA}

\author{Adilet Imambekov}
\affiliation{Department of Physics and Astronomy, Rice University, Houston, Texas 77005, USA}

\author{Dmitry A. Abanin}
\affiliation{Department of Physics, Harvard University, Cambridge MA 02138, USA}

\author{Eugene Demler}
\affiliation{Department of Physics, Harvard University, Cambridge MA 02138, USA}

\date{\today}

\begin{abstract}

Recent experimental realization of strongly imbalanced mixtures of
ultracold atoms opens new possibilities for studying impurity dynamics in
a controlled setting. In this paper, we discuss how the techniques of
atomic physics can be used to explore new regimes and manifestations of
Anderson's orthogonality catastrophe (OC), which could not be accessed
in solid state systems.

Specifically, we consider a system of impurity atoms, localized by a strong optical lattice potential, immersed
in a sea of itinerant Fermi atoms. We point out that the Ramsey interference type experiments with the impurity atoms allow one to study OC in the time domain, while  radio-frequency (RF) spectroscopy probes the OC in the frequency domain. The OC in such systems is universal for all times and not only in the long time limit and is determined fully by the scattering length and the Fermi wave vector of the itinerant fermions. We calculate the universal Ramsey response and RF absorption spectra. In addition to the standard power-law contributions, which correspond to the excitation of multiple particle-hole pairs near the Fermi surface, we identify a novel important contribution to OC that comes from exciting one extra particle from the bottom of the itinerant band. This contribution gives rise to a non-analytic feature in the RF absorption spectra, which shows a non-trivial dependence on the scattering length, and evolves into a true power-law singularity with universal exponent $1/4$ at the unitarity.

We extend our discussion to spin-echo-type experiments, showing that they probe more complicated non-equilibirum dynamics of the Fermi gas in processes in which an impurity switches between states with different interaction strength several times; such processes play an important role in the Kondo problem, but remained out of reach in the solid state systems. We show that, alternatively, the OC can be seen in the energy counting statistics of the Fermi gas following a sudden quench of the impurity state. The energy distribution function, which can be measured in time-of-flight experiments, exhibits characteristic power-law singularities at low energies.

Finally, systems in which the itinerant fermions have two or more hyperfine states provide an even richer playground for studying non-equilibrium impurity physics, allowing one to explore non-equilibrium OC and even to simulate quantum transport through nano-structures. This provides a previously missing connection between cold atomic systems and mesoscopic quantum transport.

\end{abstract}


\maketitle

\section{Introduction}
\label{intro}

Interest in nonequilibrium quantum  dynamics
has increased dramatically in the last few years following experimental
realizations of synthetic many-body systems with ensembles of ultracold
atoms~\cite{bloch_many-body_2008,ketterle_08}.
With ultracold atoms it is not only possible to prepare microscopic
systems with desired many-body Hamiltonians but, crucially for studying
dynamics, parameters of such
Hamiltonians can be changed on time scales that are much faster than
intrinsic microscopic timescales. Ultracold atomic ensembles
are also very well isolated from the environment,  so states prepared
out of equilibrium can undergo quantum evolution without relaxation or
loss of coherence~\cite{greiner_collapse_2002}. Finally a rich toolbox
of atomic physics makes it possible to provide detailed characterization
of many-body systems, which is crucial for describing complicated
transient states resulting from non-equilibrium dynamics.
Recent experimental studies addressed such questions as relaxation of
high energy metastable states~\cite{miesner_observation_1999,sadler_spontaneous_2006,strohmaier_observation_2010},
hydrodynamic expansion of strongly interacting fermions in optical
lattices~\cite{schneider_fermionic_2012}, decoherence of split condensates~\cite{shin_atom_2004,hofferberth_non-equilibrium_2007}, coherent superexchange-mediated spin dynamics~\cite{trotzky_time-resolved_2008}, spinor dynamics~\cite{Chapman,krauser_coherent_2012}, relaxation and thermalization in 1D systens~\cite{kinoshita,trotzky_probing_2012}, as well as interaction quenches in fermionic systems~\cite{Ketterle2012}.

\begin{figure}
\begin{center}
 \includegraphics[width=0.48\textwidth]{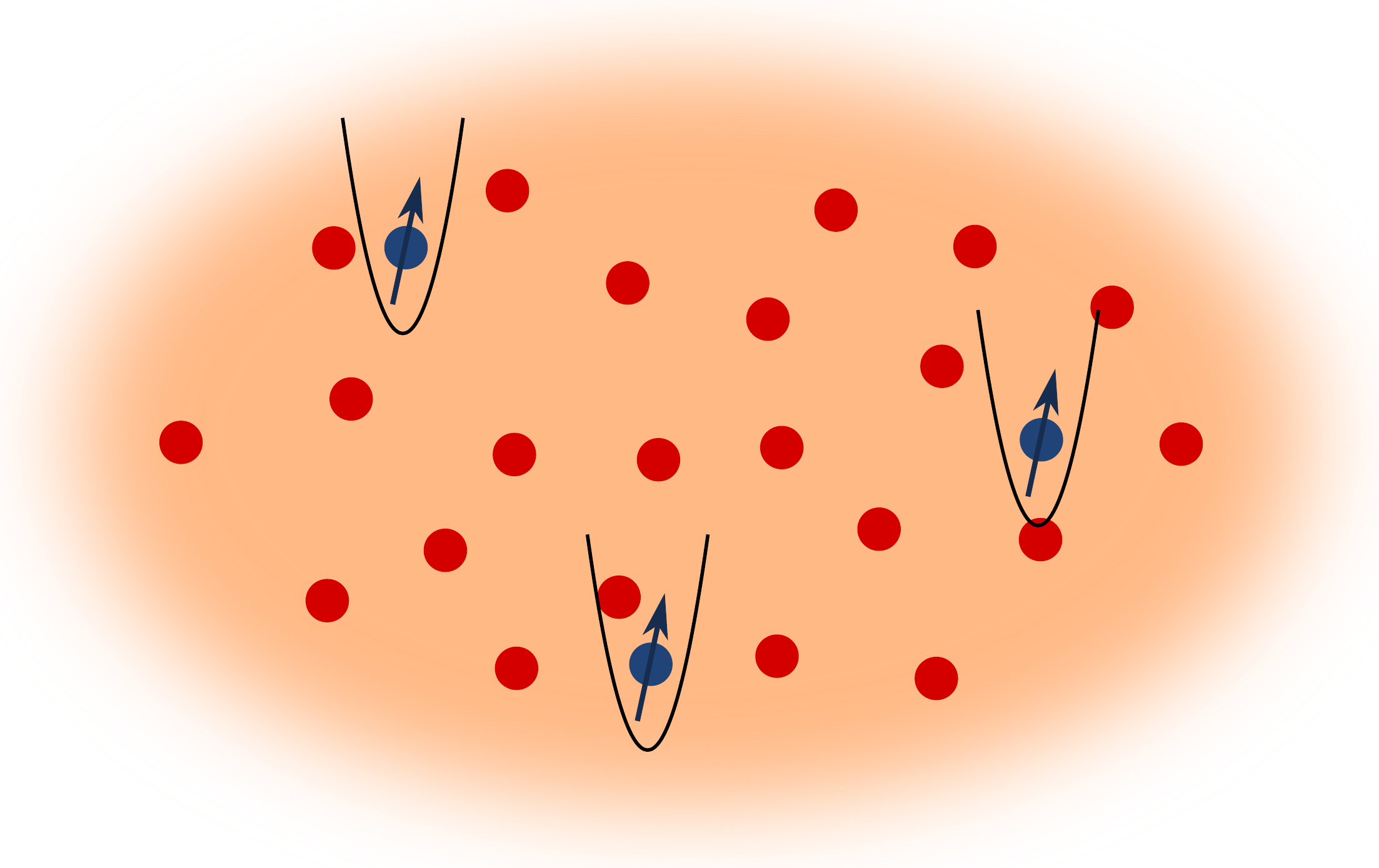}
\end{center}
\caption{\label{fig:setup} (Color online) 
Proposed experimental setup: Impurity atoms (blue dots with arrows, indicating internal states) are immersed in a sea of itinerant host fermions (red dots). The optical lattice, indicated by the black parabolas, localizes the impurity atoms without affecting itinerant
host fermions. We assume a low concentration of impurity atoms, so that we can consider
scattering on a single impurity. The scattering of fermions can be controlled
by applying magnetic fields and by manipulating the internal impurity states
with RF-pulses. The system's parameters can be controlled quickly (compared {to} the intrinsic timescales of the many-body system of host fermions), which makes this setup ideal for studying non-equilibrium impurity physics, including the new regimes of OC. }
\end{figure}

In this paper we discuss how one can use currently available experimental tools of ultracold atoms to study the problem of the orthogonality catastrophe (OC) in many-body fermionic systems. The core of this problem is understanding the response of a Fermi gas to a suddenly introduced localized impurity. This question was originally considered in the context of the X-ray absorption spectra in metals~\cite{mahan_many_2000}, where it manifests itself in the characteristic power-law threshold singularities. 

Being one of the very few known examples of a non-equilibrium solvable many-body problems, the OC also provides a conceptual framework for understanding several fundamental phenomena in solid state physics, including the Kondo effect~\cite{yuval_exact_1970}, resonant tunneling in mesoscopic structures~\cite{matveev_interaction-induced_1992,geim_fermi-edge_1994,dambrumenil_fermi_2005,abanin_fermi-edge_2005,abanin_tunable_2004,hentschel_fermi_2005}, 1D quantum physics beyond the Luttinger liquid paradigm~\cite{pustilnik_dynamic_2006,*pereira_exact_2008,*pereira_spectral_2009,imambekov_universal_2009,*imambekov_phenomenology_2009, imambekov_one-dimensional_2011}, and the motion of a heavy particle in a Fermi gas~\cite{rosch_quantum-coherent_1999}.

The basic setup that we focus on is shown in \fig{fig:setup}. We consider a quantum degenerate mixture of two types of atoms, e.g.
a Bose-Fermi mixture~\cite{schreck_quasipure_2001,ospelkaus_localization_2006,guenter_bose-fermi_2006,zaccanti_control_2006,best_role_2009,fukuhara_all-optical_2009,tey_double-degenerate_2010,wu_strongly_2011} or a Fermi-Fermi mixture~\cite{zwierlein_fermionic_2006,taglieber_quantum_2008,wille_exploring_2008,tiecke_broad_2010,liao_spin-imbalance_2010,trenkwalder_hydrodynamic_2011,hara_quantum_2011}.  We assume that one type of atoms, called an impurity atom below, has a much lower density than the other. The majority atoms, referred to as host atoms, are always taken as fermionic. Two types of atoms can have very different polarizability hence it is possible to create an optical lattice that strongly localizes impurity atoms while having very little effect  on the host fermions. The hyperfine spin states $\u, \d$ of the impurity atoms can be manipulated using RF fields, which allows to switch between weakly and strongly interacting states with respect to host fermions and thus introduce time dependent impurities in the Fermi gas~\cite{kohstall_metastability_2012}. Additionally, the strength of interactions between host atoms and a given hyperfine state of the impurities can be controlled by tuning magnetic field~\cite{schirotzek_observation_2009,nascimbene_collective_2009,navon_equation_2010,kohstall_metastability_2012,koschorreck_attractive_2012}. 
We assume a regime of very low density of impurity atoms so that scattering processes taking place on different impurities can be analyzed separately. Thus in the rest of the paper we will discuss dynamics of a single impurity interacting with a Fermi gas~\footnote{We take the energy difference between discrete levels of the impurity in a confining 
potential to be much larger than the Fermi energy of the host atoms so that we can neglect scattering of
the impurity atom into excited states.}.

The setup proposed above, with its tunability and precise control, provides a way to study new regimes and manifestations of OC, which remained out of reach in solid state systems. First, we will show that the OC is a fully universal function of the Fermi wave vector and the impurity-host-fermion scattering length; this function will be calculated below. The full solution reveals {\it new singularities} in the RF absorption spectra, which emerge away from the absorption threshold. Second, we will demonstrate that the Ramsey and spin-echo spectroscopy -- standard tools used in cold atoms -- provide a direct access to the OC, as well as to more complex non-equilibrium response functions, in the time domain. Third, manifestations of OC in the energy counting statistics will be discussed. Finally, by generalizing the setup described above, it is possible to study OC in the regime when the Fermi gas is driven out of equilibrium, as well as to simulate quantum transport. 

Given the long illustrious history of the original problem, see Refs.~\cite{mahan_excitons_1967, nozieres_singularities_1969, anderson_infrared_1967, gogolin_bosonization_2004, combescot_1971,ohtaka_theory_1990,matveev_interaction-induced_1992,geim_fermi-edge_1994,dambrumenil_fermi_2005,abanin_fermi-edge_2005,abanin_tunable_2004,hentschel_fermi_2005,mkhitaryan_fermi-edge_2011}, it is useful to start by summarizing the new aspects of OC in cold atomic systems.  

{\it Extended universality.} Experiments with ultracold atoms can demonstrate the universality of OC in a much broader sense than it was previously discussed for electron systems. The most basic quantity demonstrating OC
is the time-dependent overlap function,
\be\label{eq:overlap}
S(t)=\la\psi_0|  e^{i \hat H_i t/\hbar}  e^{-i\hat H_ft/\hbar}  |\psi_0\ra,
\ee
where $|\psi_0\ra$ is the initial ground state of the Fermi gas, and $\hat H_{i(f)}$ is the Hamiltonian before(after) the impurity is switched on. In solid state systems the universality was  understood only as a statement about the long time behavior of $S(t)$.
At long times, $t\gg \hbar/E_F$, $E_F$ being the Fermi energy, the overlap exhibits a power-law decay~\cite{nozieres_singularities_1969}, reflecting Anderson's OC~\cite{anderson_infrared_1967, gogolin_bosonization_2004}.  The power-law decay is the result of the large phase space for exciting multiple low-energy particle-hole pairs following the switching of the impurity. 

Although OC has been considered as an ``exactly solvable'' dynamical many-body problem, in solid state systems model parameters are either not known accurately or provide a crude approximation to much more complicated many-body processes. For example, a local impurity potential comes from the Coulomb potential of a hole screened by the conduction electrons.  A one particle scattering picture commonly used in the analysis of the OC~\cite{mahan_excitons_1967, nozieres_singularities_1969, anderson_infrared_1967}, which neglects many-body aspects of electron-electron interactions, is only an approximation valid for low energy scattering processes of electrons close to the Fermi energy. Furthermore, the density of states for the conduction band electrons is typically not
known and can be modified by the electron interactions. In addition, extrinsic degrees of freedom, including phonons, can play an important role in the OC dynamics leading to extra decay factors. Hence one can neither calculate the full time dependence of 
$S(t)$ nor claim its universality at all timescales. 

\begin{figure}
\begin{center}
 \includegraphics[width=3.2in]{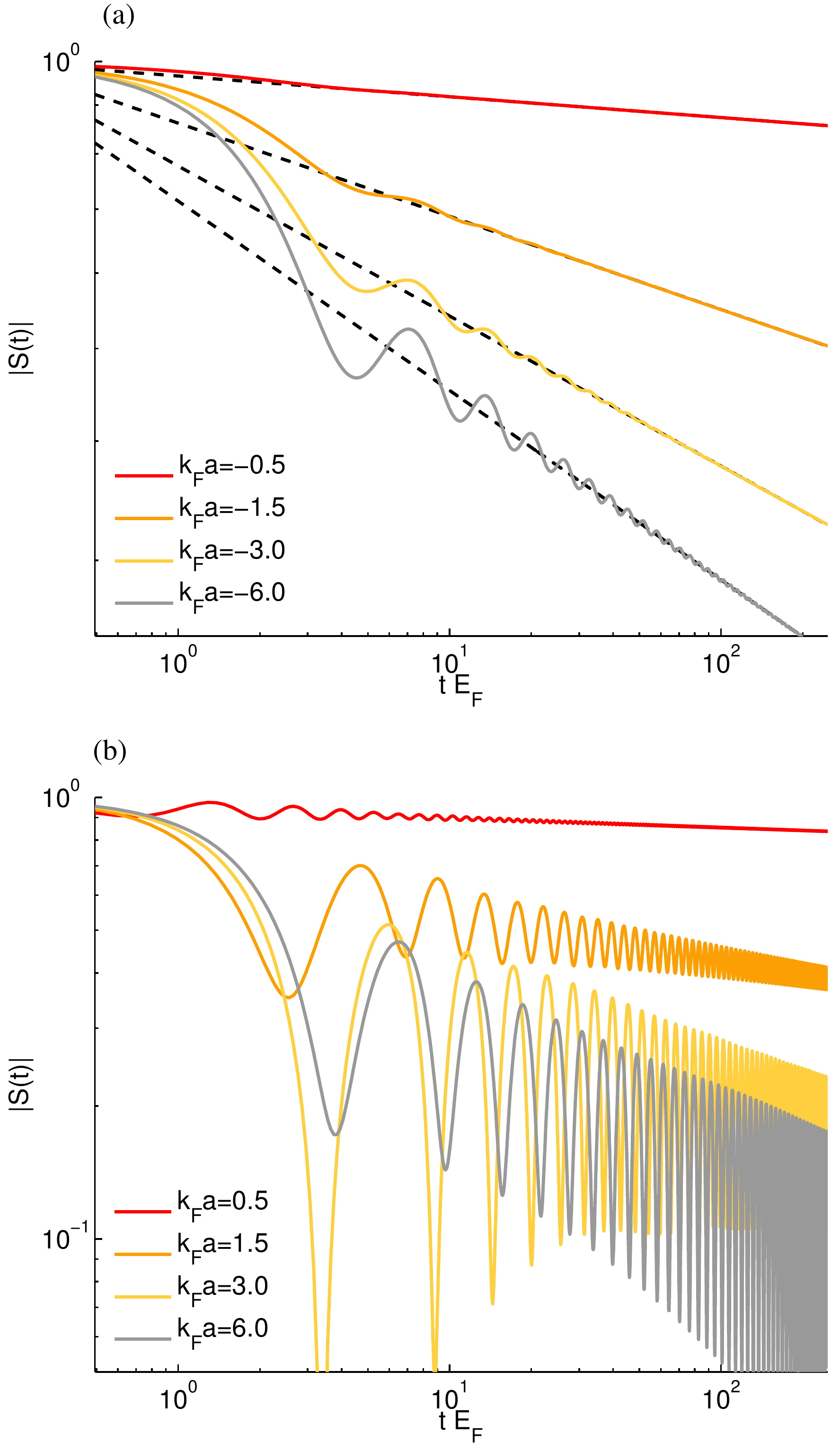}
\end{center}
\caption{\label{fig:oc} (Color online) Universal overlap function $|S(t)|$, which can be measured in the Ramsey interferometry experiment, is shown for different values of the scattering length $a$, see legend. Panel (a) shows $|S(t)|$ for the case $a<0$, when the impurity potential does not create a bound state. The dashed lines are power-law fits~(\ref{eq:overlap_fermions}) at long times. At intermediate times, oscillations coming from the bottom of the band are visible. Panel (b) shows $|S(t)|$ for the case $a>0$, when the impurity potential creates a bound state. The asymptotic behavior is described by Eq.(\ref{eq:overlap_bound}), and exhibits strong oscillations. }
\end{figure}

By contrast, in the case of ultracold atoms, one creates a gas of Fermi atoms, which are well isolated, do not interact with each other, and have a simple parabolic dispersion. The interaction of the localized impurity atoms and conduction band fermions is fully characterized by a single parameter, the scattering length $a$. In the case of wide  resonances, one finds a universal behavior of the energy dependence of the scattering amplitude 
\be\label{eq:scatt_length}
\frac{1}{f(E)}=-\frac{1}{a} + i \frac{ \sqrt{2mE}}{\hbar},
\ee 
where $E$ is the energy of the scattering atom~\footnote{It is important to point out that the scattering length $a$ characterizing the scattering of host fermions on the localized impurity is not the same as the scattering length of two free atoms.
In this paper, we take the energy difference between discrete levels of the impurity in a confining potential to be much larger than the Fermi energy of the host atoms so that we can neglect scattering of
the impurity atom into excited states; then the scattering amplitude has the canonical form (\ref{eq:scatt_length}). In principle, when the Fermi energy becomes comparable to the energy splitting between discrete levels $\omega_c$ in the parabolic potential confining the impurity, there will be an additional dependence of $f(E)$ on the $E/\omega_c$, discussed in \cite{nishida_confinement-induced_2010}; therefore, all response functions will also depend on $E_F/\omega_c$. Our discussion can be extended to find these dependencies}. Such scattering amplitude leads to the following universal behavior of the scattering phase shift on the momentum of the scattered fermion:
\be
 \delta(k)= -{\rm tan}^{-1}{ka}, 
\ee
where $-\pi/2< \delta(k)< \pi/2$. Consequently, the entire function $S(t)$ (and not only its long time asymptote) is a universal function of $k_F a$ and $E_F t$ ($k_F$ is the Fermi wave vector and $E_F$ is the Fermi energy),
\be
S(t) = S(k_F a, E_F t/\hbar).
\ee
The main result of this work is the calculation of universal functions $S(t)$, as well as more complicated time-domain response functions. Specific examples are shown in \figgg{fig:oc}{fig:spinecho}{fig:spinechoT}.

{\it Measuring orthogonality catastrophe in time domain.} In solid state systems the OC (\ref{eq:overlap}) is observed indirectly in the frequency dependence of the absorption spectrum, given by
\be\label{eq:A}
A(\omega)= \frac{1}{\pi} {\rm Re}  \int _0^\infty e^{i\omega t} S(t) \, dt.
\ee
The power-law decay of the overlap function in the time domain translates into power-law threshold singularities in the absorption spectra~\cite{gogolin_bosonization_2004, mahan_many_2000,nozieres_singularities_1969}. In ultracold atoms one can perform similar measurements of OC in the frequency domain by doing RF spectroscopy on impurity atoms. Examples of the universal absorption spectra for different values of $k_F a$ are shown in \fig{fig:spectra}. 

However, it may be more illuminating to measure $S(t)$ in time domain using Ramsey type interference, which is a {well established} experimental technique
in atomic physics. While it was originally designed for metrology applications, it has been realized recently that it can also
be used as a probe of many-body dynamics~\cite{shin_atom_2004,schumm_matter-wave_2005,gross_nonlinear_2010,kitagawa_ramsey_2010,widera_quantum_2008,gring_relaxation_2011}.  

The idea of the proposed experiment is as follows: initially, the impurity is prepared in the down-state $\d$, and the fermions are in the ground state $|\psi_0\ra$. Then, the Ramsey interferometry is performed: at time $t=0$, a $\pi/2$ pulse is applied, such that the system is driven into the superposition state
 $\frac{\d+\u}{\sqrt{2}}\otimes |\psi_0 \ra$. The two states evolve differently since the $\u$ and $\d$ states interact differently with the Fermi sea:
\be\label{eq:timeevol}
|\Psi(t)\ra= \frac{1}{\sqrt{2}} \d\otimes e^{-i \hat{H}_i t/\hbar} |\psi_0\ra + \frac{1}{\sqrt{2}} \u\otimes e^{-i \hat{H}_f t/\hbar} |\psi_0\ra.
\ee
The fermions stay undisturbed in the first state, while the impurity scattering excites multiple particle-hole pairs in the second state. Performing a second $\pi/2$ pulse after time $t$, and measuring $\la \Psi(t)| \hat S_x| \Psi(t)  \ra$ gives:
\be \label{eq:sx}
 \langle \hat S_x (t) \rangle  = \re \la \psi_0|  e^{i\hat{H}_i t/\hbar} e^{-i\hat{H}_f t/\hbar} |\psi_0 \ra = \re S(t).
\ee
In the equation above we neglected the trivial phase factor arising from the
energy difference of states $\u$ and $\d$. Thus, the Ramsey interferometry provides a direct measurement of the OC overlap~\footnote{By changing the phase $\phi$ of the closing $\pi/2$ pulse we can obtain $\re [ \, e^{i \phi} \, S(t)\, ]$}. 

The basic Ramsey type experiment described above corresponds to a local quench type dynamics in which the impurity  strength is changed once. One can also use the Hahn spin echo, as well as more complicated spin-echo-type sequences, to realize processes in which the impurity strength effectively switches between different values multiple times. As we show below the response of the Fermi gas to such processes is characterized by a power-law decay of the overlap, with an exponent {\it enhanced} compared to the case of usual OC.

The predicted faster decay of the spin-echo response goes against the atomic physics intuition; it stems from the fact that the spin-echo sequence does not ``undo" the evolution of the Fermi gas under impurity scattering, but instead, creates a state in which the impurity pseudospin and Fermi gas are strongly entangled~\footnote{We note that the idea that the interferometry of a single atom entangled with a many-body system can be employed to probe dynamics of the many-body system, has been discussed previously in a different context in Ref.~\cite{micheli_single_2004}}. From the experimental point of view spin echo experiments should have an additional advantage that they allow  to cancel slow fluctuations of the magnetic field.

{\it New universal features: bottom of the band physics.} What are the new universal characteristics that one can find in $S(t)$, as well as in the corresponding absorption spectra $A(\omega)$ at intermediate time scales? One feature that can be seen from \fig{fig:oc} is the oscillations of $S(t)$ on the timescale $\hbar E_F^{-1}$. The origin of this novel feature is the following: after the impurity is introduced, there is a class of processes in which, in addition to a large number or low-energy particle-hole pairs, an extra hole with energy $\sim E_F$ is excited near the bottom of the band. The phase space of such processes is enhanced, owing to the van Hove singularity in the density of states at the band bottom in one dimension (note that OC is essentially a 1D problem since only the $s$-wave channel is important). The combined  dynamics of the high-energy hole and the low-energy particle-hole pairs result in an additional weaker power-law contribution to the overlap function. 

This contribution to the overlap function becomes even more evident in the frequency domain, where it gives rise to a {\it cusp singularity} at the energy $E_F$ above the threshold (see Fig.~\ref{fig:spectra}). We find that,  as the Feshbach resonance is approached from the side of the negative scattering length, the cusp develops into a true singularity with a universal exponent $1/4$; for any finite value of $k_F a$, this peak is smeared, but its overall shape is described by a universal function which is discussed below.  We note that such behavior of the response function across the Feshbach resonance is rather unusual, since it shows a truly singular behavior only exactly at the resonance. For comparison, in conventional BCS-BEC crossover studies~\cite{eagles_possible_1969,*nozieres_bose_1985,*leggett_diatomic_1980,*ohashi_superfluid_2003,*ohashi_bcs-bec_2002,*chen_bcsbec_2005}, most of the measured quantities show a smooth behavior exactly at the resonance.

{\it Non-equilibrium orthogonality catastrophe.}
So far we assumed that the itinerant host fermions do not have any internal degrees of freedom.
Generalization of the setup shown in Fig.~\ref{fig:setup} to the case of a multi-component Fermi gas (with multiple internal states) allow one to realize a wide range of dynamical impurity phenomena in {\it non-equilibrium} Fermi gases.  This is an even richer class of problems that arises mesoscopics~\cite{nazarov_quantum_2009}; in particular, problems of this kind describe the quantum transport through any mesoscopic structure (e.g., a point contact), where the Fermi seas in two or more leads are kept at different chemical potentials. Mathematically, such problems can be reduced to a non-abelian Riemann-Hilbert problem~\cite{dambrumenil_fermi_2005}, the solution of which is not known in a general case. 

While usually transport is difficult to study in systems of ultracold atoms (see however~\cite{brantut_conduction_2012}), we will show that the extension of our setup provides a way of simulating the quantum transport. More generally, it allows one to study the response of the non-equilibrium Fermi gas in a controlled setting.

{\it New quantum observables and distribution functions}. OC experiments with ultracold atoms should
make it possible to study quantum variables that are not accessible in electron systems. For example, the full energy of an interacting many-body system can be measured~\cite{davis_yang-yang_2012,kinoshita_observation_2004,navon_equation_2010,ku_revealing_2012}. Moreover, it is possible to measure not 
only the average values but also fluctuations of quantum observables~\cite{folling_spatial_2005,*sanner_suppression_2010,*muller_local_2010,*perrin_hanbury_2012},
and in some cases even the full distribution functions~\cite{hofferberth_probing_2008,gring_relaxation_2011}. 

In the quantum impurity system as in \fig{fig:setup}, following the impurity potential quench the system is no longer in an energy eigenstate and the full distribution function of the total energy should also exhibit power-law type singularities (see Refs.~[\onlinecite{dambrumenil_fermi_2005,silva_statistics_2008}] and discussion below), which provides an alternative manifestation of the OC.

Full counting statistics of scattering processes should provide an intriguing connection to an extensive theoretical research in mesoscopic physics~\cite{nazarov_quantum_2009}. It is worth noting that the measurements of charge counting statistics were notoriously difficult in solid state systems~\cite{reulet_environmental_2003,*bomze_measurement_2005,*gustavsson_counting_2006}. 
Given the available experimental tools, such measurements should become possible in cold atomic systems, with an additional advantage that the full counting statistics of scattering events can be measured for fermions in specific energy windows rather than in the whole energy range.

This paper is organized as follows: In \se \ref{sec:OC} we describe our approach to OC, and present the results for the universal
overlap function $S(t)$, which can be measured by the Ramsey interferometry. The universal radio-frequency (RF) spectra $A(\omega)$ are evaluated, and their properties, singularities, as well as  the novel ``bottom of the band" feature are discussed. Extensions to the Hahn spin-echo and more complicated spin-echo type experiments, in which effectively the impurity strength changes several times, 
are studied in \se \ref{sec:secho}. The manifestations of the OC in the energy counting statistics
and generalizations to the nonequilibrium OC using multi-component host atoms are discussed
in \se \ref{sec:energy} and \se \ref{sec:quantumtransport}, respectively. Finally, we 
connect the proposed setup to existing cold-atom experiments and conclude our findings in \se \ref{sec:summary}.
\begin{figure}
\begin{center}
 \includegraphics[width=.48\textwidth]{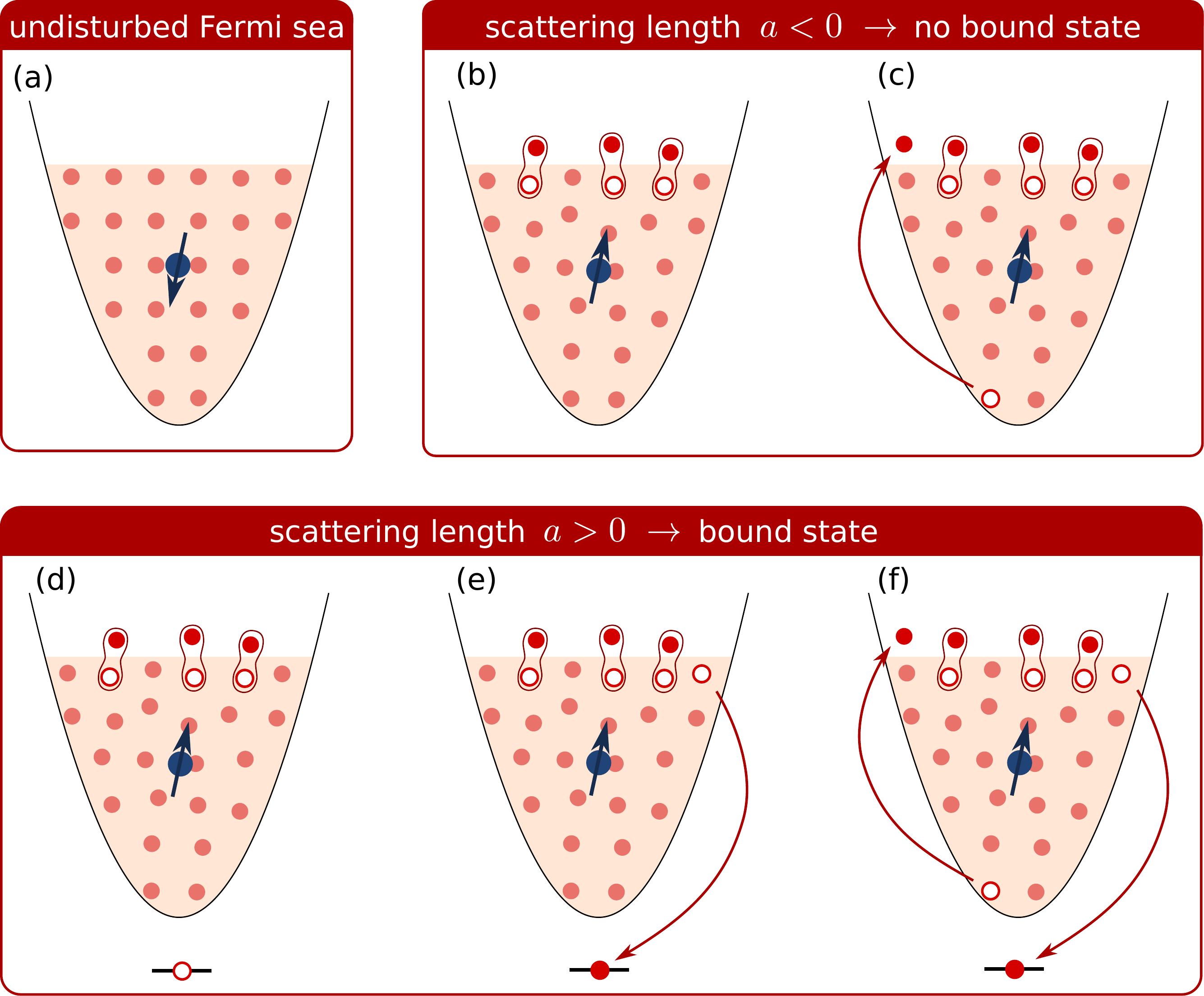}
\end{center}
\caption{\label{fig:schematic} (Color online) Schematic illustration of the intermediate excited sets of states giving power-law contributions to the overlap function, which result in the threshold singularities in the RF absorption spectrum. (a) Undisturbed Fermi sea (impurity is in the $\d$ state). (b-c) Case $a<0$ (no bound state): (b) Multiple low-energy particle-hole pairs, giving rise to the leading power-law decay (\ref{eq:overlap_fermions}) of the overlap; (c) an additional particle is promoted from the bottom of the band to the vicinity of the Fermi surface. This contribution leads to the oscillations with period $2\pi \hbar/E_F$ in the overlap function (see Fig.~\ref{fig:oc}(a)), as well as to the new cusp-like singularity in the absorption spectra (Fig.~\ref{fig:spectra}(a)). (d-f) For the case $a>0$, when the impurity potential creates a bound state, there are three important sets of states: (d) the bound state is empty and multiple low-energy excitations are created, (e) the bound state is filled, and (f) an additional particle is excited from the bottom of the band to the Fermi surface. The processes (d) and (e) lead to the behavior (\ref{eq:overlap_bound}) of the overlap function, while (f) leads to a faster decaying oscillating contribution. The three contributions result in the absorption spectra shown in Fig.~\ref{fig:spectra}(b). 
}
\end{figure} 

\section{Universal orthogonality catastrophe}
\label{sec:OC}
 
In this Section, our main goal will be to calculate the universal response functions, $S(t)$ and $A(\omega)$. 
To simplify the discussion we assume that the $\d$ state of the impurity does not interact with host fermions and 
only the $\u$ state gives rise to scattering. Generalization of our analysis to the case when both $\u$ and $\d$ states interact with the fermions is straightforward. Assuming that the impurity atom is in the rotationally symmetric ground state of the confining parabolic potential and cannot be excited to higher states (the energy of the confining potential is much larger than the Fermi energy), we only need to consider the $s$-wave scattering of host atoms on the impurity. The corresponding Hamiltonian can be written in the following form: 
\be\label{eq:hamiltonian}
\hat H=\hat{H}_0+ \u \bra{\uparrow} \otimes \hat{V}, 
\ee
where $\hat{H}_0=\sum_k \epsilon_k c_{k}^\dagger c_k$ is the Hamiltonian of free fermions, and $\hat{V}=V_0\sum_{k,q}c_k^\dagger c_q$ describes scattering. Physically, the scattering potential is characterized by the scattering length $a$. 
The Ramsey sequence described above measures the real part of the overlap function, see Eq.(\ref{eq:sx}), where $\hat H_i=\hat H_0$ and $\hat H_f=\hat H_0+\hat V$. 

There are two physically distinct cases, for which the asymptotic behavior of the overlap function (as well as the shape of the absorption spectrum) is qualitatively different~\cite{combescot_1971}: (i) $a<0$, when the impurity potential does not create a bound state, and (ii) $a>0$, when there is a bound state with energy
\be
\label{eq:Eb}
E_b=-\frac{\hbar^2}{2ma^2}.
\ee   
To gain an intuition about the behavior of the overlap function $S(t)$ in the two cases, it is convenient to rewrite it by inserting a complete set of eigenstates {$\{ \psi_\alpha \}$}  (with corresponding eigenenergies {$\{E_\alpha \}$}) of the Hamiltonian $\hat H_0+\hat V$ into Eq.(\ref{eq:overlap}):  
{\be\label{eq:S_rewritten}
S(t)=\sum_{\alpha}  |\la \psi_0| \psi_\alpha \ra|^2 e^{i (E_0-E_\alpha) t/\hbar}. 
\ee}
In \fig{fig:schematic} we illustrate the dominant intermediate states ${\ket{\psi_\alpha}}$ giving rise to the powerlaw contributions to the overlap. In the case $a<0$, the main contribution to $S(t)$ is due to the intermediate states {$|\psi_\alpha\ra$} in which multiple low-energy particle-hole pairs are created near the Fermi surface (see Fig.~\ref{fig:schematic}(b)). The large phase space of such excitations gives rise to the power-law decay of $S(t)$ at long times, see Eq.(\ref{eq:overlap_fermions}). In the case $a>0$, when there is a bound state, there are two distinct important sets of states {$|\psi_\alpha\ra$}, which involve many excitations near the Fermi surface, but differ by the bound state being either {empty or filled} (Fig.~\ref{fig:schematic}(d),(e)). The contributions of these two sets of states into $S(t)$ separate, such that at long times $S(t)$ is given by the sum of two power-laws with different exponents, see Eq.(\ref{eq:overlap_bound}). 

Below we identify another important set of intermediate states, which gives rise to an additional weaker power-law contribution to overlap. Such states involve a single hole excited from the vicinity of the band bottom, in addition to a number of low-energy particle-hole pairs (Fig.~\ref{fig:schematic}(c,f)); the contribution of these states to the overlap function is enhanced by the van Hove singularity at the bottom of 1D band. As we shall see below, this new contribution results in an unusual kink feature in the absorption spectrum, which at the unitarity {($|k_F a|= \infty$)} develops into a full power-law singularity with a universal exponent $1/4$. 

\subsection{Method}

 Our analysis of the universal OC is based on the representation of the response functions in terms of functional determinants~\cite{combescot_1971,klich_03,dambrumenil_fermi_2005,abanin_fermi-edge_2005}. We are interested in evaluating many-body averages of products of exponents of operators quadratic in creation/annihilation operators [see \eqq{eq:hamiltonian}{eq:sx}]. Such many-body quantities can be reduced to functional determinants in the space of {\it single-particle} orbitals. The overlap $S(t)$, in particular, can be written in the following form~\cite{klich_03,dambrumenil_fermi_2005,abanin_fermi-edge_2005}: 
\be\label{eq:overlap_single2}
S(t)={\rm det}\left( 1-\hat{n}+\hat{R}(t)\hat{n} \right), \,\, \hat{R}(t)=e^{i\hat h_0t/\hbar}e^{-i(\hat h_0+\hat v)t/\hbar}, 
\ee
where $\hat{n}$ is the occupation number operator, $\hat{n}|\epsilon\ra=n(\epsilon)|\epsilon\ra$; at finite temperature, $n(\epsilon)$ is given by the Fermi distribution, $n(\epsilon)=\frac{1}{\exp((\epsilon-\mu)/T)+1}$. The operators $\hat h_0, \, \hat h_0+\hat v$ are the  single-particle Hamiltonians in the absence and presence of impurity, respectively. 

While previous work~\cite{dambrumenil_fermi_2005,abanin_fermi-edge_2005,abanin_tunable_2004,protopopov_many-particle_2011} concentrated on analyzing the asymptotic behavior (at times $t\gg \hbar/E_F$) of the functional determinant of the type (\ref{eq:overlap_single2}), here our goal is to find the response functions at all times. Therefore, we numerically evaluate the functional determinants. We consider the case of a finite system confined in a ball. This allows us to view the operator in Eq.(\ref{eq:overlap_single2}) as a finite-dimensional matrix for the case of zero temperature, when $n(\epsilon)$ is non-zero only for a finite number of states under the Fermi level. It turns out that the matrix elements of the matrix $\hat R(t)$ can be evaluated to high precision by appropriately truncating the (infinite dimensional) matrix $e^{-i(\hat h_0+\hat v) t}$.  Taking the finite size scaling, we obtain the universal behavior of the overlap in the continuum limit. The details of our discretization and truncation procedures can be found in Appendix \ref{app:numerics}. An advantage of our approach is that it can be easily generalized to other geometries, including the experimentally important case of the harmonic trap.

The structure of the single-particle Hilbert space depends on the sign of the scattering length: when $a<0$, the eigenstates of the Hamiltonian $\hat h_0+\hat v$ are scattering states; when $a>0$, the spectrum should be supplemented by the bound state. This translates into a different asymptotic behavior of the overlap function $S(t)$ and the absorption function $A(\omega)$, as discussed below.

 \subsection{Universal overlap functions}

\begin{figure}
\begin{center}
\includegraphics[width=3.2in]{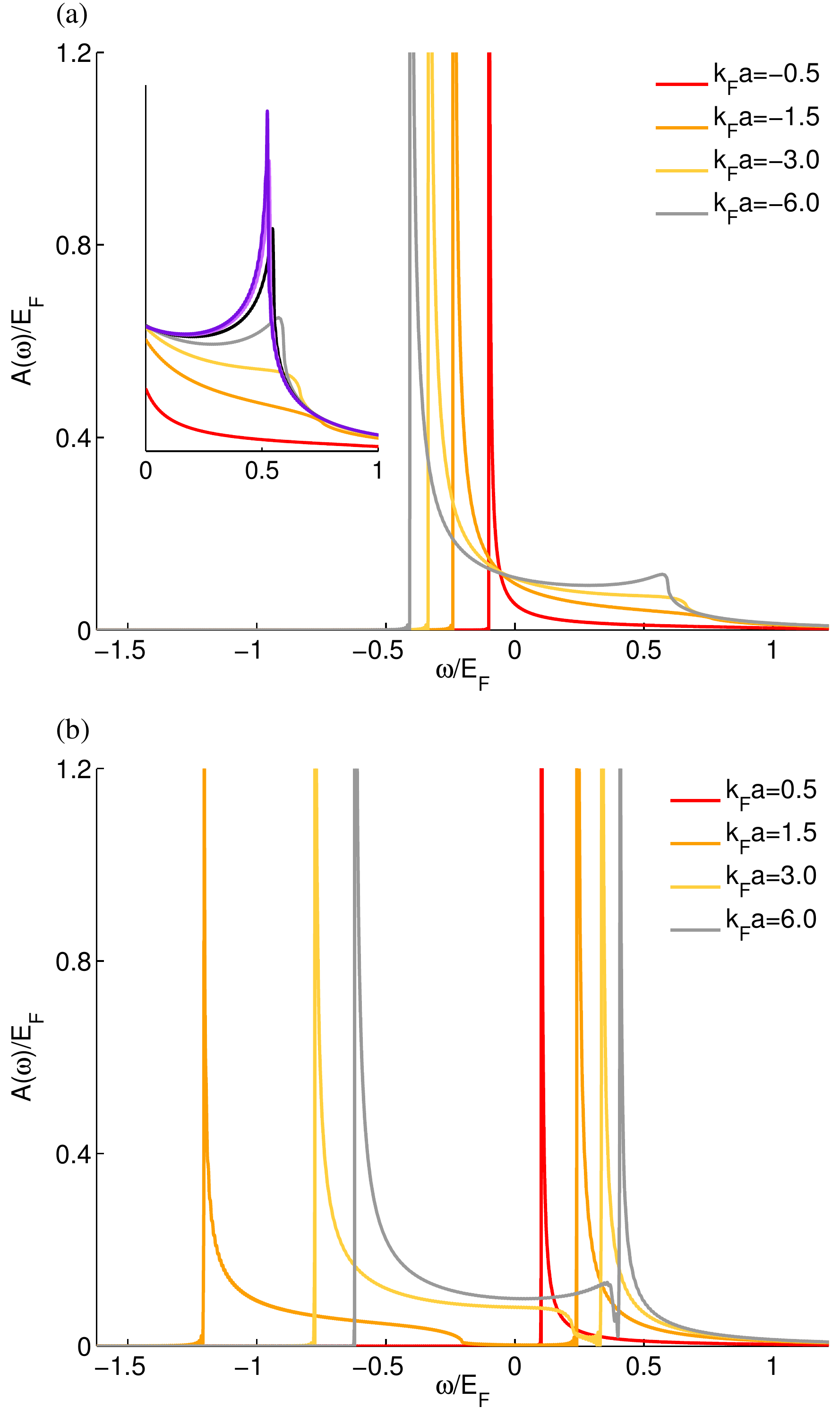}
\end{center}
\caption{\label{fig:spectra} (Color online) RF absorption spectra for different values of scattering length $a$. (a) When $a<0$ no bound state is present and the spectrum exhibits one power-law edge singularity at $\hbar \omega_0=\Delta E$, due to the processes illustrated in Fig.~\ref{fig:schematic}(b). In addition, a weak non-analyticity at 
$\hbar \omega_1=\Delta E+E_F$ is found, attributed to the excitations from the bottom of the band,
illustrated in Fig.~\ref{fig:schematic}(c). This non-analyticity is magnified and plotted for additional large values of $k_F a = -12,-18,-24$ in the inset in (a). At unitarity, the non-analyticity transforms into a true divergence with exponent 1/4, described by a universal result~(\ref{eq:universal_response}). (b) When $a>0$, the impurity creates a bound state. The spectra in this case exhibit two threshold singularities, which are offset by an energy $|E_F-E_b|$, as well as an additional cusp-like singularity the origin of which is illustrated in Fig.~\ref{fig:schematic}(f).}
\end{figure} 

The absolute value of the resulting universal function $S(t)$ in the two cases is shown in Fig.~\ref{fig:oc}. (We note that the quantity $\re S(t)$, measured in the Ramsey experiment, is quickly oscillating with a period set by the energy difference between two hyperfine states as well as by the difference of the ground state energies of the Hamiltonians $\hat H_0$ and $\hat H_0+\hat V$; when plotting the Ramsey response in Fig.~\ref{fig:oc}, we have ignored this trivial phase and illustrated the Ramsey amplitude $|S(t)|$.)   

\begin{figure}
\begin{center}
 \includegraphics[width=0.48\textwidth]{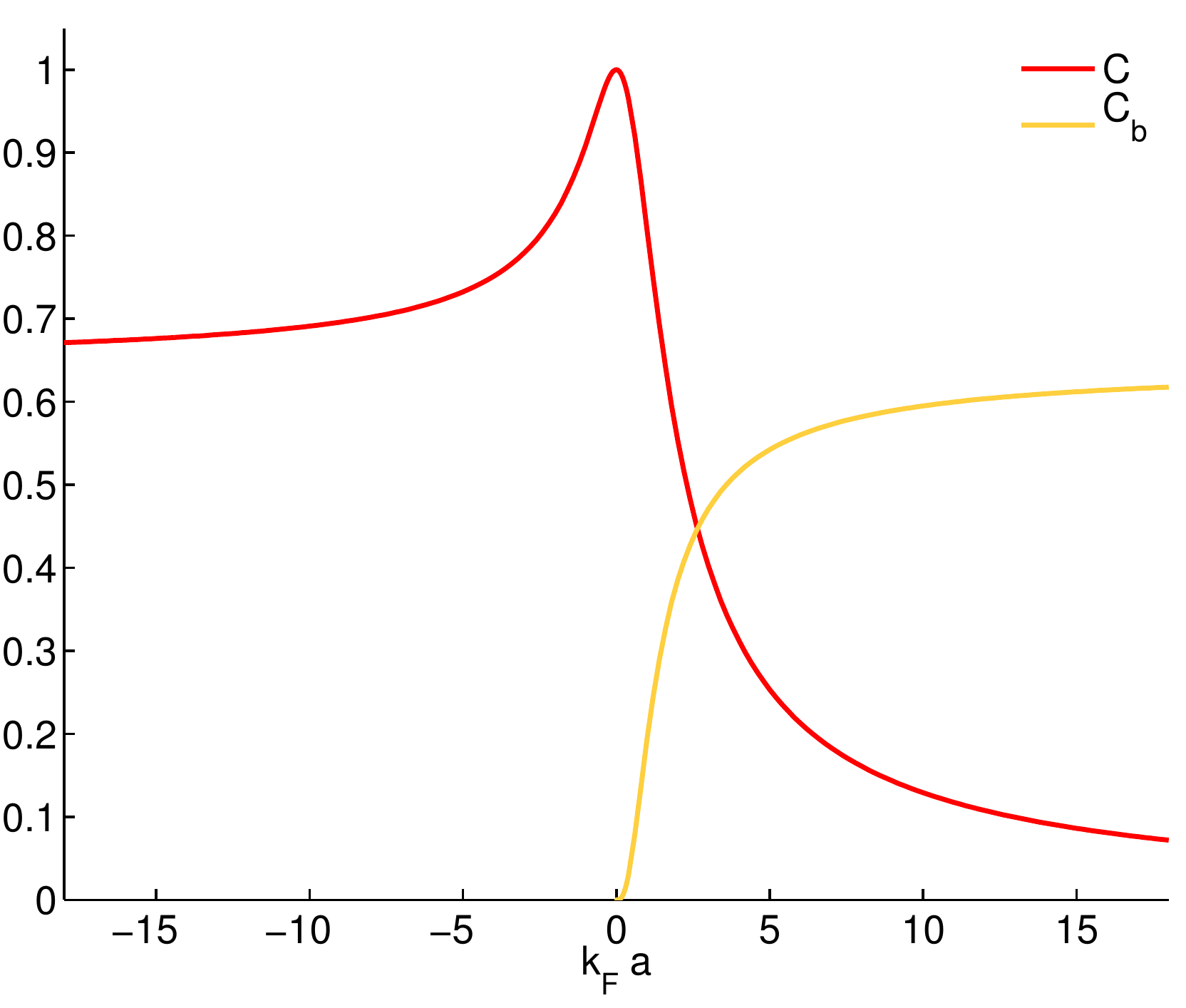}
\end{center}
\caption{\label{fig:prefactors} (Color online) The prefactors that appear in the asymptotic behavior of the overlap function, as well as in the RF response near the  thresholds at $\omega_0, \omega_b$, see {Eqs.~(\ref{eq:overlap_fermions}),(\ref{eq:overlap_bound}),(\ref{eq:oc_response}), and (\ref{eq:oc_boundstate_response})}. The prefactors were obtained from the asymptotic form of the numerically evaluated overlap functions.
}
\end{figure}

In the case $a<0$, see \figc{fig:oc}{a}, the long-time asymptotic behavior at $t\gg \hbar/E_F$ agrees with the analytic result~\cite{nozieres_singularities_1969}
\bea
\label{eq:overlap_fermions}
S(t) \approx C e^{-i\Delta E t/\hbar}\left(\frac{1}{i E_F t/\hbar+0}\right)^{\alpha}, \;\;  \alpha=\frac{\delta_F^2}{\pi^2},
\eea
where 
\be
\delta_F =  -{\rm tan}^{-1}(k_F a)
\ee
is the scattering phase shift at the Fermi wavevector, 
\be
\label{eq:energy_shift}
\Delta E= -\int_0^{E_F} \frac{dE}{\pi} \delta(\sqrt{2mE})
\ee
is the energy renormalization of the Fermi sea due to the impurity level~\cite{affleck_boundary_1997}. The power law in \eq{eq:overlap_fermions} is a manifestation of the ``shake-up" process after the sudden switching of the impurity potential, which creates multiple low-energy particle-hole excitations near the Fermi surface, see Fig.~\ref{fig:schematic}(b). The dependence of the prefactor $C$ on $k_F a$, calculated from fitting the numerical results at long times with formula (\ref{eq:overlap_fermions}), is illustrated in Fig.~\ref{fig:prefactors}. The overlap function $S(t)$ exhibits oscillations with period $2\pi \hbar/E_F$ (see Fig.~\ref{fig:oc}), which are due to the processes in which an extra particle is excited from the bottom of the band to the Fermi level (see Fig.~\ref{fig:schematic}(c)).

For the case $a>0$, see \figc{fig:oc}{b}, the overlap function $S(t)$ has two main contributions, coming from the intermediate states where the bound state is either filled or empty. This alters the asymptotic behavior, which is now represented by the sum of two power laws~\cite{combescot_1971}: 
{\allowdisplaybreaks \bea\label{eq:overlap_bound}
S(t)\approx & & Ce^{-i\Delta E t/\hbar} \left(\frac{1}{iE_Ft/\hbar+0}\right)^{\alpha}+\nonumber\\
& &C_b e^{-i(\Delta E -E_F+E_b)t/\hbar} \left(\frac{1}{iE_F t/\hbar + 0}\right)^{\alpha_b},\nonumber\\
& & \,\, \alpha_b=(1+\delta_F/\pi)^2, 
\eea}
where $C$ and $C_b$ are $k_F a$ dependent numerical constants. The strong oscillations with period $2\pi\hbar/(E_F-E_b)$, predicted by the above formula, are evident in Fig.~\ref{fig:oc}(b). Our approach allows the numerical calculation of the prefactors $C$ and $C_b$, yielding the result shown in Fig.~\ref{fig:prefactors}~\footnote{We note that these prefactors can also be investigated from  following the method of Ref.~\cite{shashi_nonuniversal_2011,*shashi_exact_2012}. The details of such procedure which leads to fully analytical results will be published elsewhere~\cite{prefactor_paper}}. In addition, the overlap function exhibits faster-decaying oscillations with period $2\pi \hbar/E_F$, which correspond to the ``bottom of the band'' contribution (see Fig.\ref{fig:schematic}(f)); in Fig.~\ref{fig:oc}(b), these are not visible because they are masked by the stronger oscillations with period $2\pi\hbar/(E_F-E_b)$.

\subsection{Universal radio-frequency spectra}
\label{sec:rf}

We now use the above results for the overlap function to evaluate the RF spectra. Numerically calculating $S(t)$ for a very large interval of $t$ values, necessary for a precise Fourier transform, is computationally challenging. We circumvent this difficulty as follows: we choose a cut-off $t_*\gg \hbar/E_F$; at times $t< t_*$, function $S(t)$ is evaluated numerically, while at $t>t_*$, we match $S(t)$ to its asymptotic power-law behavior given by \eq{eq:overlap_fermions} for $a<0$ and \eq{eq:overlap_bound} for $a>0$. 
Calculating the real part of the Fourier transform of the resulting function, we obtain the RF spectra. 

The behavior of $A(\omega)$ for negative and positive scattering length is qualitatively different (see Fig.~\ref{fig:spectra}). The absorption spectrum for $a<0$ exhibits one edge singularity, with an exponent $1-\alpha$ that follows from Eq.(\ref{eq:overlap_fermions}).  For $a>0$, the asymptotic behavior (\ref{eq:overlap_bound})  gives rise to a double-threshold absorption spectrum; due to the presence of the bound state the spectrum is characterized by two singularities at energies $\hbar \omega_b=\Delta E-E_F+E_b$ and $\hbar \omega_0=\Delta E.$ 
The first threshold corresponds to filling the bound state following absorption, and the second to leaving it empty~\cite{combescot_1971}. As follows from the definition of $A(\omega)$, \eq{eq:A}, and from the asymptotic form (\ref{eq:overlap_bound}), the two threshold singularities are characterized by different exponents, $1-\alpha_b$ and $1-\alpha$, respectively. The excitation processes which correspond  to these singularities are illustrated in Figs. {\ref{fig:schematic}(e) and \ref{fig:schematic}(d)}. The response near these thresholds can be obtained using formulas (\ref{eq:overlap_fermions},\ref{eq:overlap_bound}), which gives:
\bea
\label{eq:oc_response}
S(\omega-\omega_0) &\approx& \frac{2\pi C \theta(\omega-\omega_0)}{\Gamma(\alpha)(E_F/\hbar)^\alpha}|\omega-\omega_0|^{\alpha-1},\nn
& &\hspace{2.5 cm} \hbar\omega_0 = \Delta E,\\
\label{eq:oc_boundstate_response}
S(\omega-\omega_b) &\approx& \frac{2\pi C_b \theta(\omega-\omega_b)}{\Gamma(\alpha_b)(E_F/\hbar)^{\alpha_b}}|\omega-\omega_b|^{\alpha_b-1},\nn
& &\hspace{1 cm} \hbar\omega_b = \Delta E - E_F + E_b.
\eea

The most interesting new feature of the universal spectra shown in Fig.~\ref{fig:spectra} is the non-analyticity  of $A(\omega)$ at frequency $\hbar \omega_1=\Delta E+E_F$ {for $a<0$ and $\hbar \omega_1=\Delta E+E_b$ for $a>0$}; the physical origin of this feature was already discussed above.   
As the unitarity is approached, the non-analytic feature becomes more prominent. This phenomenon, as well as the full structure of $A(\omega)$  at $k_F|a|\gg1$ can be understood as the result of a non-trivial interplay between two-body physics that involves the impurity and the hole near the bottom of the band, and the dynamics of multiple particle-hole excitations created at the Fermi surface. The possibility of a non-trivial interplay between many-body and few-body physics is a unique feature of ultracold atom physics, and has attracted significant theoretical~\cite{tan_energetics_2008,*tan_large_2008,*tan_generalized_2008, braaten_universal_2008,*braaten_exact_2008,*haussmann_spectral_2009,*combescot_particle_2009,*werner_number_2009,*schneider_universal_2010,*braaten_universal_2011,*zhang_universal_2009} and experimental~\cite{kuhnle_universal_2010,*stewart_verification_2010,*wild_measurements_2012} interest recently.

We have developed an analytic theory of this new feature; here, we briefly summarize the essence of our approach and main results, providing the full results in Appendix \ref{sec:appendix}; the details of the solution will be provided in Ref.~[\onlinecite{prefactor_paper}]. The idea is that, in the time domain, the contributions from the Fermi surface excitations and from the single hole at the bottom of the band to the determinant (\ref{eq:overlap_single2}) factorize. The former contribution is given by the power law which can be obtained within the functional determinant approach, while the latter can be evaluated exactly within the two-body theory~\cite{prefactor_paper}. 

In the frequency domain, the result can be written as a convolution of the two terms describing these two processes (see Appendix \ref{sec:appendix}); this gives rise to a cusp-like singularity at an energy {$\Delta E+E_F$ for $a<0$ and $\Delta E+E_b$ for $a>0$}. Away from the unitarity, this singularity is smeared on the energy scale $\sim \hbar^2/2ma^2$ (for either sign of the scattering length $a$); the origin of this smearing lies in the dynamics of the hole.  

Remarkably, right at the unitarity, {$|k_F a| = \infty$, $\hbar^2/2ma^2$} vanishes, and a full non-smeared power-law singularity with the universal exponent $1/4$ develops; this singularity is asymmetric, {and is described by:}
\bea
\label{eq:universal_response}
A(\omega)&\approx& \frac{1.74 |\omega-\omega_1|^{-1/4}}{(E_F/\hbar)^{3/4}}\times\nn
& &\bigg[ \theta(\omega-\omega_1)\frac{\Gamma(1/2)}{\Gamma(3/4)} + \theta(\omega_1-\omega)\frac{\Gamma(1/4)}{\Gamma(1/2)}\bigg].
\eea
for {$|k_Fa|=\infty$} and $\hbar|\omega-\omega_1|\ll E_F.$

When $a<0$, this peak gets smeared out  at energies of the order of $\hbar^2/2ma^2$ from its maximal value; the evolution of the peak depending on the value of $k_F a$ is illustrated in the inset to Fig.~\ref{fig:spectra}(a). 
 
For the case $a>0$, the true bound state with energy $E_b$ ``pinches off" from the bottom of the band and leads to a threshold with an exponent $3/4$. Thus for large but finite  $k_Fa>0$ the universal form of $A(\omega)$ near $\omega_1$ has a characteristic double peak structure, as is seen in Fig.~\ref{fig:spectra}(b) for $k_Fa= 6.0$. For increasing interaction parameter $k_F a$ the bottom of the band feature approaches the singularity at $\omega_0$ where the bound state is empty, as the energy difference is only of the order of $E_b$ {which tends to zero when the unitarity is approached}.

The universal contribution due to the bottom of the band is a unique feature of ultracold spinless fermions which has no analog in conventional solid state systems. Indeed, the universal behavior coming from the excitations in the vicinity of the Fermi surfaces is ubiquitous in solid state systems, and is controlled by Fermi liquid theory \cite{nozieres_theory_1997}. On the other hand, away from the Fermi surface fermionic quasiparticles are not well defined in 3D interacting systems, and thus bottom of the band contributions to the orthogonality catastrophe cannot be probed in solid state systems. In contrast, for spinless cold atoms, the bottom of the band contributions are well defined, since the 
interactions between atoms in the $s$-wave channel are absent. Thus, fermionic excitations are well defined for all energies, including the vicinity of the band bottom.

\begin{figure}
\begin{center}
 \includegraphics[width=0.48\textwidth]{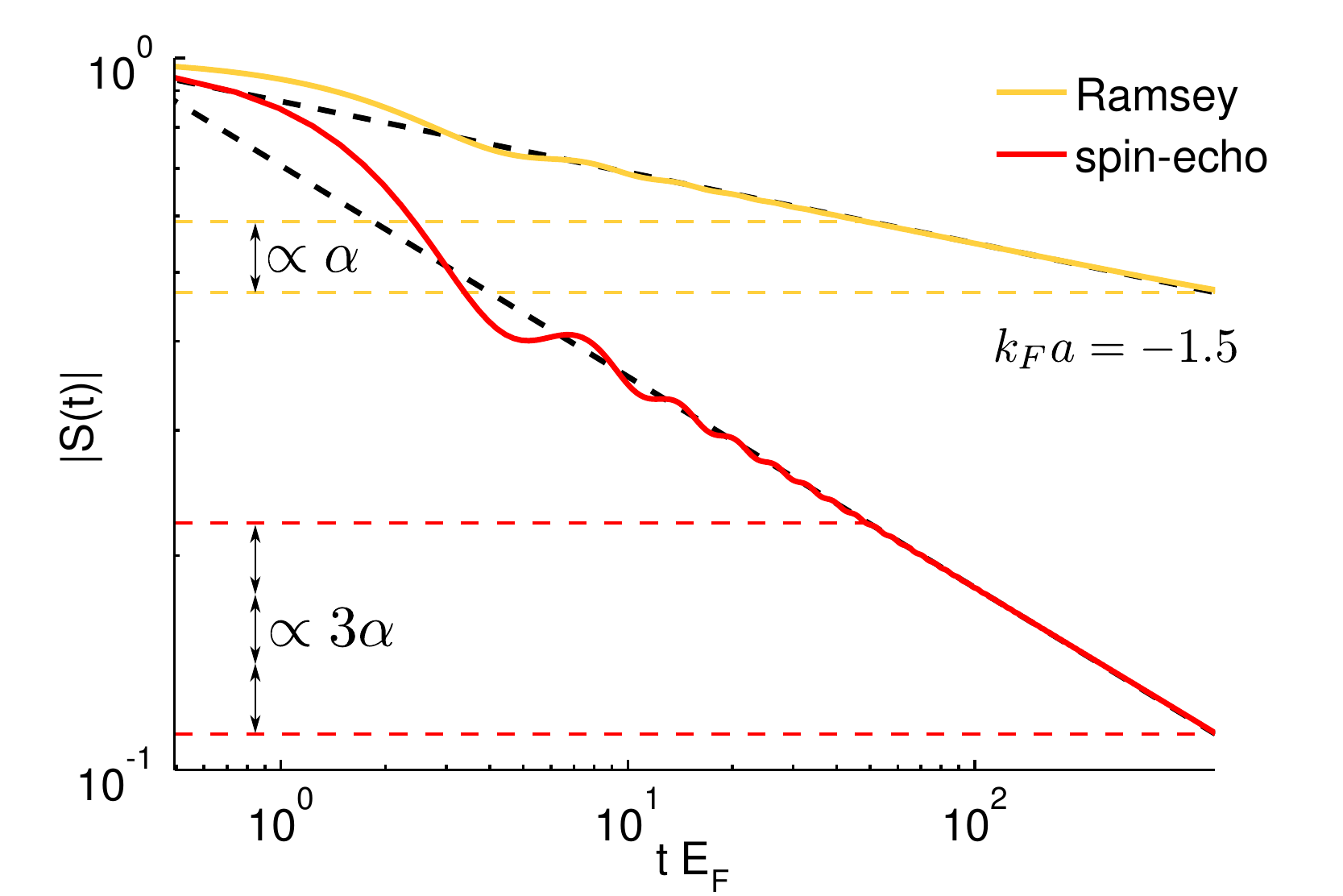}
\end{center}
\caption{\label{fig:spinecho} (Color online) The spin-echo response (\ref{eq:spin_echo_Sx2}) of the Fermi gas. At long times, it is characterized by a power-law decay (\ref{eq:spin-echo}) with an exponent three times larger than that of the standard OC. 
}
\end{figure} 

\begin{figure*}
\begin{center}
 \includegraphics[width=0.98\textwidth]{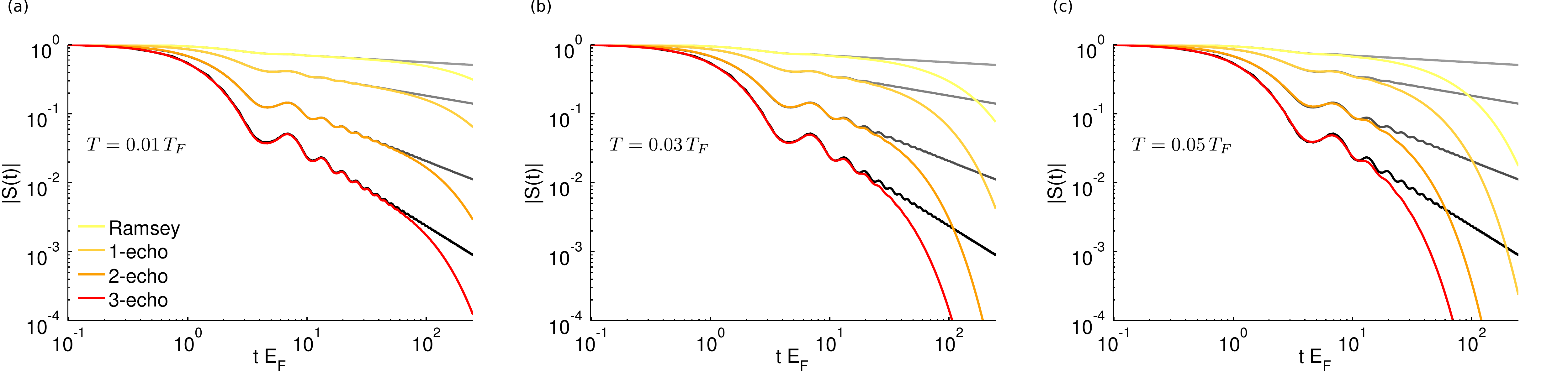}
\end{center}
\caption{\label{fig:spinechoT} (Color online) The generalized spin-echo response (\ref{eq:spin_echo_Sx_n}) of the Fermi gas for $n=1,2,3$ compared to the Ramsey interference. At long times, the generalized spin-echo response is characterized by a power-law decay (\ref{eq:sxn}) with an exponent $(4n-1)$ times larger than that of the standard OC. The gray lines correspond to $T=0$, while the colored lines are for finite temperature \fc{a} $T=0.01 \,T_F$, \fc{b} $T=0.03 \,T_F$, and \fc{c} $T=0.05 \,T_F$. The scattering length is $k_F a= -1.5$ in all cases. The power-law decay of the spin-echo response gives way to a faster exponential decay at times $t\gtrsim \hbar/T$.  
}
\end{figure*}

\section{Spin echo response}
\label{sec:secho}

Now we proceed with discussing spin-echo-type sequences, which allow one to measure the response of the Fermi gas to a process in which the internal impurity states switch multiple times. The spin-echo sequences have an important advantage compared to the Ramsey interferometry in that they are not sensitive to the slowly fluctuating magnetic fields typically present in experiments. 

The main effect of the fluctuating fields is to introduce an energy difference $\Delta \varepsilon$ between the $\u$ and $\d$ states. Generally, this would affect the outcome of the Ramsey experiment: owing to the different phase picked up by the $\u$ and $\d$ states, the measured value of $\la \hat S_x\ra$ is given by
$$
\la \hat S_x(t)\ra=\frac{S(t)e^{i \Delta\varepsilon t/(2\hbar)}+S^*(t)e^{-i\Delta  \varepsilon t/(2\hbar)}}{2}. 
$$
The phase factor $e^{i\Delta  \varepsilon t/(2\hbar)}$ is different from shot to shot; thus, averaging over different shots gives rise to an additional decay of $\la S_x (t)\ra$. This extra decay would complicate the observation of the OC. Similarly to quantum optics experiments, the undesired contribution of the fluctuating magnetic fields can be eliminated in the spin-echo experiment. 

We first consider the Hahn spin-echo protocol: initially, the system is prepared in the state $\d \otimes |\psi_0\ra$; at time $-t$ a $\pi/2$ pulse is applied, followed by the $\pi$ pulse at $t=0$. Finally, at time $t$ another $\pi/2$ pulse is applied. Similar to the case of the Ramsey experiment, we can show that $S_x$ measured after such a sequence is given by: 
\be\label{eq:spin_echo_Sx2}
\la \hat S_{x,1} (t)\ra ={\rm Re}[S_1(t)], 
S_1(t)=\la \psi_0| \hat U_2^{-1} \hat U_1 | \psi_0\ra, 
\ee
where 
\[
\hat U_1= e^{-i\hat H_0t/\hbar}   e^{-i(\hat H_0+\hat V)t/\hbar}, \,\,\hat U_2=e^{-i(\hat H_0+\hat V)t/\hbar} e^{-i\hat H_0t/\hbar},
\]
are the operators describing the evolution of the Fermi gas state that was initially coupled to the $\u$ and $\d$ states of the impurity. 
This correlation function describes non-trivial Fermi gas dynamics in a process where the impurity switches between different states several times. Similar response functions arise in the analysis of the Kondo problem~\cite{yuval_exact_1970}. 
To understand the behavior of the spin-echo response, we represent it as a functional determinant given by Eq.(\ref{eq:overlap_single2}) with $\hat R=\hat u_2^{-1}\hat u_1$ ($\hat u_1,\hat u_2$ being the single-particle analogues of operators $\hat U_1,\hat U_2$). 
As can be shown analytically using the method of Ref.~[\onlinecite{dambrumenil_fermi_2005}], the asymptotic behavior of such a determinant at long time $t\gg \hbar/E_F$ is given by a power-law,
\be\label{eq:spin-echo}
S_{1}(t)\propto t^{-3\delta_F^2/\pi^2}, 
\ee
with an exponent three times larger than that of OC. 
The universal behavior of the spin-echo response for arbitrary times, calculated numerically using the determinant approach, is compared to the standard OC probed with Ramsey interference  in Fig.~\ref{fig:spinecho}.

The predicted faster decay of the spin-echo response is somewhat unexpected: the intuition from quantum optics would suggest that the spin-echo response generally eliminates the broadening due to slowly fluctuating environment, and therefore should be characterized by the slower decay compared to the Ramsey interference. This intuition, however, relies on the assumption that the environment only affects the relative phase of the two hyperfine states; thus, the operators describing the effect of the environment on the two pseudospin states commute (and therefore the spin echo sequence can ``undo'' the effect of the environment). The above discussion illustrates that this assumption breaks down for the case when the Fermi gas plays the role of environment: the reason is that the operators $e^{-i(\hat H_0+\hat V)t}$ and $e^{-i\hat H_0t}$ no longer commute, and therefore $\hat U_1\neq \hat U_2$; this results in the non-trivial form of the spin-echo response (\ref{eq:spin_echo_Sx2}) and its faster decay.

The generalized spin-echo sequence, in which $(2n-1)$ $\pi$ pulses are applied at equal time intervals $t$ (the Hahn spin-echo corresponds to $n=1$), allows measurements of even more interesting response functions. The corresponding overlap function is given by:
\be\label{eq:spin_echo_Sx_n}
\la \hat S_{x,n}(t) \ra ={\rm Re}[S_n(t)], \; S_n(t)=\la \psi_0| \hat U_2^{-n}  \hat U_1^n  | \psi_0\ra.
\ee
The long-time asymptotic behavior of the $n$th response, calculated similar to the case of the Hahn spin-echo, is characterized by the power-law decay 
\be\label{eq:sxn}
S_{n}(t)\propto t^{-\alpha_n \delta_F^2/\pi^2}, \,\, \alpha_n=4n-1. 
\ee 
Thus, by increasing the number of pulses $n$, the exponent can be strongly enhanced. This should facilitate the observation of the power-law decay in the spin-echo experiments. The universal spin-echo responses for different values of $n$, calculated numerically, are illustrated in Fig.~\ref{fig:spinechoT} along with Ramsey interference results. The asymptotic form at large times agrees with the analytic formula (\ref{eq:sxn}). 

Experiments are always done at low, but finite temperature; therefore, it is important to understand the effect of non-zero temperature $T$ on the spin-echo response. Within our approach, the finite temperature response is found by calculating the corresponding determinants with the distribution functions $n(\epsilon)=\frac{1}{\exp((\epsilon-\mu)/T)+1}$. The result, illustrated in Fig.~\ref{fig:spinechoT}, shows that the finite temperature does not affect the power-law behavior of the response at relatively short times $t\lesssim \hbar/T$; however, at longer times, $t\gtrsim \hbar/T$, the responses are characterized by a faster, exponential decay. It is evident from Fig.~\ref{fig:spinechoT} that two most interesting features of the spin-echo response -- the power-law decay and the oscillations with period $2\pi \hbar/E_F$ -- survive up to experimentally accessible temperature $T=0.05 E_F$, and therefore they should be observable with current experimental means.

\section{Manifestations of the OC in energy counting statistics}
\label{sec:energy}

So far, we have considered two ways of observing OC---the RF spectroscopy (energy domain), and the Ramsey, as well as spin echo sequences (time domain). Both methods are based on studying the impurity properties. However, cold atomic system also allow direct measurements of the Fermi gas properties (e.g., measuring the occupation numbers of states with different momenta in the time-of-flight experiments). Can one see traces of OC by measuring properties of the Fermi gas following sudden change of an impurity potential? 

An obvious candidate quantity is the time-dependent density profile of the Fermi gas. We have studied the time-dependent density profile $\rho(t,r)$, finding that it does not carry signatures of the OC. This is because the overlap function involves a complicated sum of the $n$-particle-hole pair correlation functions, while the density operator can create at most one particle-hole pair. Thus, we are forced to consider the {\it fluctuations} rather than averages of physical quantities.  

We note that the OC is formally related to the distribution function of the {\it fluctuations of the total energy of the Fermi gas} following an impurity quench. For simplicity, let us consider a situation where the impurity is initially in the interacting state $\u$, such that fermions are in the appropriate ground state $|\psi_0'\ra$; at some time, 
the impurity is suddenly flipped to the non-interacting state $\d$.  Then, the characteristic function of the total energy distribution (with new Hamiltonian) is given by: 
\[
\chi (\lambda)=\la \psi_0' | e^{i\lambda \hat H_0}|\psi_0'\ra. 
\]
Comparing with Eq.(\ref{eq:overlap}), we see that up to a phase factor, the characteristic function is identical to the overlap function, with parameter $\lambda$ playing the role of time. Thus, $\chi(\lambda)$ exhibits power-law behavior at large $\lambda$:
\be\label{eq:chi_long}
\chi (\lambda)\propto  \lambda^{-\delta_F^2/\pi^2}.   
\ee

The power-law asymptotic behavior (\ref{eq:chi_long}) gives rise to the universal power-law behavior of the probability distribution $P(E)$ at low energies $E\ll E_F$: 
\be\label{eq:prob}
P(E)=\frac{1}{2\pi}\int e^{-i\lambda E} \chi(\lambda) d\lambda \propto E^{\delta_F^2/\pi^2-1}. 
\ee
The unusual distribution function can be directly measured in the time-of-flight experiments (the total energy is given simply by the kinetic energy since in the final state the fermions are not interacting with the impurity). The power-law singularity (\ref{eq:prob}) in the energy fluctuations, similar to that found in the context of spin systems by Silva~\cite{silva_statistics_2008}, provides a new experimentally accessible manifestation of OC. 

\section{Non-equilibrium OC and quantum transport}
\label{sec:quantumtransport}

The orthogonality catastrophe is modified qualitatively when the Fermi gas is driven out of equilibrium~\cite{dambrumenil_fermi_2005,abanin_fermi-edge_2005}. In solid state systems, such a situation occurs naturally when an impurity is coupled to two 1D leads with different chemical potentials. The impurity state controls the transmission between the two leads. From the mathematical point of view, the problem of non-equilibrium OC can be reduced to  a non-abelian Riemann-Hilbert problem~\cite{dambrumenil_fermi_2005}, which has not been solved in the general case. Progress has been made in the limit of long times, where the OC is characterized by the combination of a power-law decay with new exponents and weak exponential damping~\cite{abanin_fermi-edge_2005,dambrumenil_fermi_2005}. 

We now argue that it is possible to study non-equilibrium OC in cold atomic gases by generalizing the setup proposed above to the case of a multi-component Fermi gas. For simplicity, let us consider a two-component gas, with pseudospin species $|u\ra$ and $|d\ra$. Our goal is to realize a situation in which two components are at different chemical potentials (playing the role of two leads), and an impurity that can scatter the fermions between two species. The first condition can be achieved by preparing an imbalanced Fermi gas, with different Fermi energies of two components, $E_{F}^u=E_0$, $E_{F}^d=E_0+\Delta \mu$. The second condition is more difficult to attain: an impurity's scattering matrix is diagonal in $|u\ra$, $|d\ra$ basis, with phase shifts at the Fermi energy given by $\delta_{u,d}$. In the solid state analogy, this corresponds to an impurity always fully reflecting electrons; effectively, this brings us back to two copies of the equilibrium OC. 

To overcome this difficulty, we consider an application of $\pi/2$ pulse to the pseudo spin of the host fermions. After that, the $|1\ra= \frac{|u\ra+|d\ra}{\sqrt{2}}$  and  $|2\ra=\frac{|u\ra-|d\ra}{\sqrt{2}}$ states of fermions will be at different chemical potentials. Crucially, the impurity's scattering matrix is non-diagonal in $|1\ra$, $|2\ra$ basis (it is obtained by a rotation of the S-matrix in the $|u\ra$, $|d\ra$ basis by matrix $\frac{1+i\sigma_y}{\sqrt{2}}$). The $|1\ra, |2\ra$ species play the role of the electrons in the left and right leads in the mesoscopic experiment, with the impurity being characterized by a non-trivial scattering matrix. This is exactly the situation needed for realizing non-equilibrium OC. Performing the Ramsey or spin-echo experiments on the impurity pseudospin then allows one to study the response of the non-equilibrium Fermi gas. 

Looking beyond OC, the analogy between the setup we just considered and a quantum point contact (QPC) with an impurity controlling the transmission through the QPC suggests the possibility of studying the full counting statistics of charge transfer. Although theoretically charge counting statistics has played an important role~\cite{levitov_electron_1996}, its experimental studies in mesoscopic systems have been quite  limited~\cite{reulet_environmental_2003,*bomze_measurement_2005,*gustavsson_counting_2006}. Simulating quantum transport in cold atom experiments is also attractive because time-of-flight experiments allow energy-resolved measurements; this can be used to study correlations between number of transmitted particles at different energies.

\section{Conclusions}
\label{sec:summary}

In conclusion, we studied universal OC in cold atomic systems, and discussed the manifestations of OC in the Ramsey and spin-echo interferometry, RF spectroscopy, as well as in energy counting statistics. Beyond the equilibrium OC, we have proposed a set of experiments which probe the dynamics and transport in nonequilibrium Fermi gases. This provides a useful connection between the cold atom physics and mesoscopic physics.

The basic ingredients required for the experimental implementation of our proposal are the following: (a) a quantum degenerate Bose-Fermi~\cite{schreck_quasipure_2001,ospelkaus_localization_2006,guenter_bose-fermi_2006,zaccanti_control_2006,best_role_2009,fukuhara_all-optical_2009,tey_double-degenerate_2010,wu_strongly_2011} or Fermi-Fermi~\cite{zwierlein_fermionic_2006,taglieber_quantum_2008,wille_exploring_2008,tiecke_broad_2010,liao_spin-imbalance_2010,trenkwalder_hydrodynamic_2011,hara_quantum_2011} mixture, (b) the ability to trap one type of atom by a strong optical lattice potential, (c) the control of the interaction strength between the impurity and host atom via changing the impurity hyperfine state, and (d) the ability to achieve temperatures that are low enough to observe the OC. 

We emphasize that all four requirements are achievable with currently available techniques, thus we expect that our proposal can be implemented in the near future. Below we describe some relevant experiments, which, we hope, will help to identify the systems which are most suitable for studying OC. 
\begin{enumerate}
\item[(a)] Various quantum degenerate mixtures have been realized by multiple groups~\cite{schreck_quasipure_2001,ospelkaus_localization_2006,guenter_bose-fermi_2006,zaccanti_control_2006,best_role_2009,fukuhara_all-optical_2009,tey_double-degenerate_2010,wu_strongly_2011,zwierlein_fermionic_2006,taglieber_quantum_2008,wille_exploring_2008,tiecke_broad_2010,liao_spin-imbalance_2010,trenkwalder_hydrodynamic_2011,hara_quantum_2011}.
\item[(b)] Recently, the localization of minority atoms by an optical lattice was demonstrated with an imbalanced Bose-Bose mixture of ${}^{87}$Rb and ${}^{41}$K atoms~\cite{catani_quantum_2012}. The localization of the impurity atoms at length scales of roughly $10\%$ of the Fermi wavelength has been
achieved at typical densities; thus, the impurity can be treated as point-like, and one can neglect its excitations to the excited states of the trapping potentials, as in our analysis above. 

\item[(c)] RF pulses have been used to switch between the hyperfine states of the impurity atom, which interact differently with the host fermions~\cite{kohstall_metastability_2012}. This should enable the Ramsey and spin-echo type experiments that reveal OC. 
Furthermore, experiments with strongly imbalanced
Fermi-Fermi mixtures, which addressed the polaron dynamics (mobile impurities)~\cite{schirotzek_observation_2009,nascimbene_collective_2009,
navon_equation_2010,kohstall_metastability_2012,koschorreck_attractive_2012} demonstrated that the impurity-host atom
interactions can be tuned in the full interaction range, from strong attractive to strong repulsive regime. This should allow the exploration of different regimes of the OC, discussed above. 

\item[(d)] In current experiments, temperatures as low as a few percent of the Fermi
temperature can be achieved~\cite{partridge_deformation_2006,navon_equation_2010,ku_revealing_2012}. As follows from our analysis above, this should be sufficient to observe both the universal power law
decay as well as the ``bottom of the band'' oscillations with period $2\pi \hbar/E_F$.
In RF absorption spectra these oscillations result in a cusp-like
singularity at the Fermi energy $E_F$. 

\end{enumerate}

It is also worth noting that, for very short times our results
may be relevant to very heavy mobile impurities (mass 
much larger than that of the host atoms). OC will manifest itself, e.g., in the RF spectroscopy experiments, which are commonly used to probe mixtures of cold atoms. A suitable mixtures with large mass ratio are  ${}^{40}$K/${}^{41}$K immersed in ${}^6$Li~\cite{taglieber_quantum_2008,wille_exploring_2008,tiecke_broad_2010,wu_strongly_2011,trenkwalder_hydrodynamic_2011} and ${}^{173}$Yb/${}^{174}$Yb immersed in ${}^6$Li~\cite{hara_quantum_2011}. 

An alternative experimental route to accessing the non-equilibrium response of the Fermi gas to a suddenly introduced impurity is to create a local scattering potential by a narrow laser beam. This has been demonstrated in {recent experiments~\cite{weitenberg_single-spin_2011,desbuquois_superfluid_2012}}. While in such a setup one cannot perform the RF absorption or Ramsey interference
experiments, an observation of OC should be possible through energy 
counting statistics, see \se \ref{sec:energy}. Furthermore, time-of-flight experiments would reveal the full distribution functions of scattering processes, providing new information about the non-equilibrium state of fermions.

Finally, we note that the ideas presented above are not limited to the case of free fermions; one particularly interesting extension concerns the case of 1D {\it interacting} fermions in an optical lattice. OC in Luttinger liquids (LLs) was predicted to exhibit strong deviations from the non-interacting case, showing new power laws depending on the strength of interactions, universal power-law asymptotic behavior, as well as new scaling laws at intermediate times~\cite{gogolin_local_1993,*prokofev_fermi-edge_1994,*meden_orthogonality_1998}. To access the possibility of observing these phenomena in a 1D optical lattices, we have carried out a numerical study~\cite{unpublished} of interacting spinless fermions in 1D lattice; we found that the modifications of OC in LL can be observed in finite optical lattices with only $\sim 100$ atoms, which put them within the reach of current experiments.

\section{Acknowledgements}
We thank R. Grimm, C. Salomon, I. Zapata, and especially M. Zwierlein for many inspiring discussions. We acknowledge support from Harvard-MIT CUA, the NSF Grants No. DMR-07-05472 and No. DMR-1049082, the DARPA OLE program,
AFOSR Quantum Simulation MURI, AFOSR MURI on Ultracold Molecules, 
the Austrian Science Fund (FWF) under Project No. P18551-N16 (M.K.), Austrian Marshall Plan Foundation (M.K.), the Welch Foundation, Grant No. C-1739, the A.P. Sloan Foundation (A. S. and A. I.), as well as by LANL Oppenheimer Fellowship (Y.N.). 

\appendix

\section{Details on the numerical procedure}
\label{app:numerics}

In order to obtain the universal overlap function $S(t)$, we evaluate determinants
of type \eqw{eq:overlap_single2} numerically. To this end, we consider a finite system
confined in a sphere of radius $R$ whose eigenstates are in the absence and in the 
presence of the impurity
\begin{align}
\label{eq:single_p_wf}
 \psi_n(r) &= \sqrt{\frac{2}{R}} \sin (k_n r) \, ,\quad k_n R=\pi n \nn
 \psi'_n(r) &= A_n \sqrt{\frac{2}{R}} \sin (k'_n r+\delta_n) \, ,\quad k_n' R + \delta_n = \pi n \;,
\end{align}
respectively, where $A_n=1/\sqrt{1+\frac{\sin2\delta_n}{2k_n'R}}$ and $\delta_n=-\tan^{-1}(k_n' a)$. For $a>0$ there is also a bound state which must be treated separately. 
In principle the evolution of the determinants requires even at zero temperature
multiplications of infinite dimensional matrices. However, truncating the infinite number
of intermediate states still allows to evaluate the determinants with very high accuracy, since the OC is determined
by low energy physics. In order to obtain the universal overlap functions $S(t)$
from the finite system we take a finite size scaling by keeping the density constant
while taking the system size to infinity. Importantly, in a finite size system
the overlap function $S(t)$ does not decay to zero but rather exhibits revivals
after sufficiently long times characterized by the Fermi velocity and the system size
{---an additional aspect that should be observable in experiments.}

We evaluate the universal RF-spectra $A(\omega)$ from the Fourier transform
of the overlap function $S(t)$. The RF-spectra $A(\omega)$ exhibits 
power law decays and band edges, which renders 
a numerical evaluation of the Fourier transform extremely challenging, since $S(t)$
has to be known for extremely long times.  To circumvent this problem,
we calculate $S(t)$ exactly from \eq{eq:overlap_single2} up to a certain
time $t<t^* \sim 500\hbar E_F^{-1}$ and then continue $S(t)$ to longer times
by fitting it to its asymptotic form \eqw{eq:overlap_fermions} and \eqw{eq:overlap_bound}, 
respectively. Even though, with that we have access to $S(t)$ for long times,
the Fourier transform is still finite dimensional, which gives rise to 
wild oscillations at the edges of $A(\omega)$, known as Gibbs phenomenon. 
We reduce the Gibbs phenomenon by applying a Lanczos filter to the 
Fourier coefficients. With this procedure, we produce the universal RF-spectra shown 
in \fig{fig:spectra}. Importantly, the positions of the band edges in the
spectra match exactly the outcome of \eqq{eq:oc_response}{eq:oc_boundstate_response}.

\vspace{0.5cm}

\section{Universal radio-frequency spectra}
\label{sec:appendix}

Here we provide the analytic formulas that describe the new feature in the absorption spectra $A(\omega)$ at frequency $\hbar \omega_1=\Delta E+E_F$ {for $a<0$ and $\hbar \omega_1=\Delta E+E_b$ for $a>0$},  the origin of which is illustrated in Fig.~\ref{fig:schematic}(c) ($a<0$) and (f) ($a>0$). The configurations which lead to this non-analyticity are given by the excitation of a fermion from the bottom of the band to the Fermi surface. 
The numerical results in Fig.~\ref{fig:spectra} indicate that the non-analytic feature develops into a stronger peak as unitarity is approached. This phenomenon, as well as the full structure of $A(\omega)$  at $k_F|a|\gg1$ can be understood as a simple interplay between two-body physics happening near the bottom of the band, and physics of multiple particle-hole excitations being created at the Fermi surface.  

In the time domain, the contributions from the Fermi surface excitations and the dynamics of the hole at the bottom of the band factorize. The former contributions manifest as 
the usual power laws with unitary phase shift, while the latter can be simply evaluated within two-body theory, see~\cite{prefactor_paper} for details. In the frequency domain,
the result, written as a convolution of the two terms corresponding to these two processes, reads:
\begin{widetext}
\bea
\label{eq:full_shadow_response}
A(\omega)\approx\frac{1.74\pi }{\Gamma(1/4)(E_F/\hbar)^{1/4}}  \int_{-\infty}^{\infty}\frac{d\varepsilon}{2\pi}\theta\left(\omega-\omega_1 + \frac{\varepsilon}{\hbar}\right)\left(\omega-\omega_1 + \frac{\varepsilon}{\hbar}\right)^{-3/4} F(\varepsilon)+reg.,
\label{Sconv}
\eea
where $F(\varepsilon)$ is the probability to excite a hole with energy $\varepsilon$ measured from the bottom of the band,
\bea
F(\varepsilon) = \theta(\varepsilon)\frac{2\sqrt{\hbar\varepsilon/E_F}}{\frac{\hbar^2}{2ma^2}+\varepsilon}+\theta(k_Fa) \frac{4\pi\delta(\varepsilon-E_b) }{k_Fa}.
\label{Fdef}
\eea
\end{widetext}
Numerical prefactor is evaluated similar to $C,$ and  function $F(\varepsilon)$ accounts for the existence of the bound state on the repulsive side.
Terms denoted as $reg.$ account for a regular contribution  at  $\omega_1$ which is not singular at  large $k_F|a|.$

Although for {$0<\varepsilon\ll \hbar^2/2ma^2$} function  $F(\varepsilon)$ behaves as $\propto \sqrt{\varepsilon,}$ and leads to a weak non-analyticity, for {$\hbar^2/2ma^2\ll\varepsilon,$} it behaves as $\propto 1/\sqrt{\varepsilon.}$  Right at unitarity  the scale $E_b$ disappears, and one obtains in $A(\omega)$ the divergence with the universal exponent $1/4$ and a universal shoulder ratio, see Eq.(\ref{eq:universal_response}), which is valid 
for {$|k_Fa|=\infty$} and $\hbar|\omega-\omega_1|\ll E_F.$
 For large but finite negative $k_Fa,$ this peak gets smeared out  at energies of the order of {$\hbar^2/2ma^2$} from its maximal value, as is illustrated in the inset to Fig.~\ref{fig:spectra}(a). On the repulsive side, the true bound state with energy $E_b$
``pinches off" from the bottom of the band and leads to a threshold with an exponent $3/4$. Thus for large but finite  $k_Fa>0$ the universal form of $A(\omega)$ near $\omega_1$ has a characteristic double peak structure, as is seen in Fig.~\ref{fig:spectra}(b) for $k_Fa= 6.0$.

%

\end{document}